\documentclass{article}

\usepackage{arxiv}

\usepackage[utf8]{inputenc} 
\usepackage[T1]{fontenc}   
\usepackage{hyperref}       
\usepackage{url}            
\usepackage{booktabs}       
\usepackage{amsfonts}     
\usepackage{nicefrac}       
\usepackage{microtype}     
\usepackage{lipsum}
\usepackage{changepage}
\usepackage{graphicx}
\usepackage{amsmath}
\usepackage{amsfonts}
\usepackage{amssymb}
\usepackage{makeidx}
\usepackage{algorithm,algorithmic}
\usepackage[switch]{lineno}
\usepackage{xcolor}
\usepackage{bm}
\usepackage[justification=centering]{caption}
\usepackage{amsfonts}
\usepackage[bbgreekl]{mathbbol}
\DeclareSymbolFontAlphabet{\mathbbm}{bbold}
\DeclareSymbolFontAlphabet{\mathbb}{AMSb}

\usepackage{amsthm}
\newtheorem{theorem}{Theorem}
\newtheorem{problem}{Problem}
\newtheorem{definition}{Definition}
\newtheorem{remark}{Remark}
\newtheorem{lemma}{Lemma}

\usepackage{array, booktabs, makecell}

\title{State-Space Neural Network with Ordered Variance \\ for Model Order Determination}

\author{
  Midhun T. Augustine,  Mani Bhushan, and Sharad Bhartiya\\Automation Lab, 
  Department of Chemical Engineering\\
  Indian Institute of Technology Bombay, India\\
  \texttt{\{30004946,mbhushan,sharad\_bhartiya\}@iitb.ac.in}  \\
  \vspace{.01cm}\\
  Date of initial version: 14 - 06 - 2024\\ 
   Date of current version: 30 - 05 - 2026\\\vspace{.01cm} }

\begin{document}
\maketitle

\begin{abstract}
This paper addresses the  problem of identifying a nonlinear state-space model, along with an adequate model order, from a given input-output training dataset. To this end, a novel framework, termed state-space neural network with ordered variance (SSNNO), is proposed. In SSNNO, the state variables are ordered according to their variances computed using the training data. This ordering is achieved by introducing a variance-regularization term into the loss function used for SSNNO training and it facilitates a distinction between significant states, which exhibit high variance from the other residual states with near-zero variance. The number of significant states is indicative of a suitable model order. The variance-regularization mechanism is designed to minimize the number of significant state variables, thereby promoting a minimal order of the identified state-space model without significantly compromising its prediction accuracy. 
A systematic procedure is then introduced to obtain a reduced-order state-space model from the trained SSNNO, yielding a reduced-order SSNNO (R-SSNNO).
The existence of an SSNNO with variance-ordered states, based solely on input-output data, as well as an upper bound on its output prediction error, are formally established. A practical and robust method is proposed for ensuring variance-ordered states in an SSNNO, even when the network is trained using local optimization algorithms.
The effectiveness of the proposed method for identification of nonlinear state space models is demonstrated through simulation studies on a nonlinear continuous stirred-tank reactor process. The identified model is further used for state estimation and prediction in a model predictive control implementation.

\end{abstract}


\section{Introduction}
\par Identifying dynamic models from data finds applications in control, estimation, process monitoring, and economics,  among other areas \cite{bLL99,bSY14,bMB12}. The initial works were related to identifying input-output linear dynamic models such as transfer functions \cite{bLL99}, autoregressive exogenous models \cite{bSB13}, as well as nonlinear descriptions such as radial basis functions  \cite{bSJ01}, kernel methods \cite{bGP14},  feedforward neural network (NN) and recurrent NN (RNN) models \cite{bKS90,bLCK20}. These methods typically rely on training data derived from available sensor measurements and known exogenous inputs. 
However, a key limitation of input-output models is their inability to capture the internal dynamics of the system. 
In contrast, state-space models not only represent the relationship between inputs and outputs but also incorporate the internal state of the system, providing a more comprehensive description of system dynamics. Consequently, state-space models are widely adopted in modern control methodologies, such as  {model predictive control} (MPC) \cite{bFB17}. 
This has led to various approaches for identifying linear state-space models: prediction error methods (PEMs)
     \cite{bRD17}, realization theory-based approaches \cite{bBR66}, and subspace projection methods \cite{bBB96}. %All three identification approaches for linear state-space models are mature with standard software available \cite{bLL99}.   
     However,  linear state-space models provide only a local approximation of the nonlinear system dynamics and are inadequate for applications that require a nonlinear description over a wide operating envelope. This has led to nonlinear state-space identification approaches: probabilistic \cite{bTS11}, {kernel method \cite{bMS20}, subspace identification \cite{bSA23}}, autoencoder \cite{bDM21}, and {state-space neural network} (SSNN)  \cite{bJS95}, among others. Most of the above approaches for nonlinear system identification can be classified as nonlinear extensions of PEMs. 
      \par Among PEMs, SSNNs, a form of RNNs that learn nonlinear state-space models, have recently gained popularity, at least partly due to the availability of software for training deep neural networks and methods for stability analysis\cite{bJZ98, bKK18}. %The motivation for NN-based approaches such as  SSNN comes from the {universal approximation theorems}, which prove the ability  of NNs to approximate continuous nonlinear functions with arbitrary accuracy \cite{bGC89,bSP21}.
     %The fact that the stability of an SSNN can be investigated by standard tools of system analysis \cite{bJZ98, bKK18} makes it an ideal candidate model structure for use in model-based estimation and control methods. 
  Despite these advantages, SSNNs are plagued with similar issues as encountered in PEMs, namely convergence to local minima, unknown initial condition, and unknown state dimension, leading to over-parameterization of the SSNN model. Convergence of SSNN to local minima during training has been explored in \cite{bAR20,bMS21}. Estimating the initial state of SSNNs has also been investigated in \cite{bDM21,bGB23}. However, the issue of unknown model order remains largely unaddressed. 
In \cite{bKK18,bMS21,bGB23}, the dimension of the state vector is assumed to be known {a priori}. However, the order of a real system is generally unknown, and this fact imposes a serious limitation on the identification of SSNN models since the selection of the model order by trial and error requires considerable training and testing. Although model order determination methods are proposed for autoencoder \cite{bDM21} and kernel-based nonlinear models \cite{bSA23}, their extension to SSNNs is not explored in the literature.
\par This motivates the proposed {state-space neural network with ordered variance} (SSNNO) approach, which makes use of a novel idea of using a variance-regularization term in the loss function of SSNN that leads to simultaneously identifying an adequate model order (or state dimension) as well as the nonlinear state-space model. %Once the variances of the identified states are ordered, then the model order is estimated as the number of states that exhibit significant variance over the training data and whose value becomes apparent during the identification step.
To the best of the authors' knowledge, this is the first work that incorporates variance-ordering of state variables in SSNN. 
The proposed SSNNO is inspired by a previous work \cite{bMPMS24} that uses the idea of variance-ordering of latent variables, resulting in an autoencoder with ordered variance (AEO). 
The major contribution of the current work lies in proposing a systematic approach for estimating the model order with SSNNO. We establish the existence of such an SSNNO whose state variances are ordered, and provide a practical method to obtain such a model. An upper bound on the output prediction error of the constructed SSNNO is also derived. 
Further, a model order reduction step is incorporated to obtain a reduced order SSNNO (R-SSNNO) model. 
%The applicability of R-SSNNO is demonstrated in the contexts of control: model predictive control (MPC), and state estimation:  extended Kalman filter (EKF) using a simulation case study.
% \par The rest of the paper is organized as follows. Section 2  reviews relevant concepts from SSNNs and the problem of model order determination. The proposed SSNNO approach is presented in Section 3. This section also presents the determination of the reduced model order, followed by steps to obtain an R-SSNNO model. Section 4 presents theoretical results on variance ordering and a bound for output prediction error with SSNNO.
%  Section 5 
%  illustrates the numerical implementation of SSNNO for the identification of a nonlinear CSTR system and the comparison of results with  SSNN, followed by data-driven MPC using the SSNNO model. Conclusions and future directions are discussed in Section 6.

\par \textit{Notations:} % The $n$-dimensional Euclidean space is denoted by  $\mathbb{R}^{n}$, and the space of $m \times n$ real matrices by $\mathbb{R}^{m \times n}$. 
The sample mean vector and sample covariance matrix of the vector $\mathbf{x}$ over the dataset $\{\mathbf{x}_{0},\mathbf{x}_{1},\dots,\mathbf{x}_{{N}-1} \}$ are defined as $\bar{\mathbf{x}}=\frac{1}{{N}}\sum_{k=0}^{{N}-1}\mathbf{x}_{k}$ and $ \mathbf{V}_{\mathbf{x}}=\frac{1}{{N}-1}\sum_{k=0}^{{N}-1}(\mathbf{x}_{k}-\bar{\mathbf{x}})(\mathbf{x}_{k}-\bar{\mathbf{x}})^{\top}$. 
The Euclidean norm of vector $\mathbf{x}\in \mathbb{R}^{n}$ is defined as ${\parallel \mathbf{x} \parallel}_{2}=\sqrt{\sum_{i=1}^{n} x_{i}^{2} } $
 and  the Frobenius norm of  matrix $\mathbf{A}\in \mathbb{R}^{m \times n}$ is 
${\parallel \mathbf{A} \parallel}_{F}=\sqrt{\sum_{i=1}^{m}\sum_{j=1}^{n}a_{ij}^{2}} = \sqrt{\operatorname{trace}(\mathbf{A}^{\top}\mathbf{A})}.$ $\mathbf{I}_{n}$ represents the identity matrix of size $n\times n$.

\section{State-Space Neural Network with Ordered Variance (SSNNO)}  
\begin{figure}[h!]
 	\begin{center}
 		\includegraphics [scale=0.7] {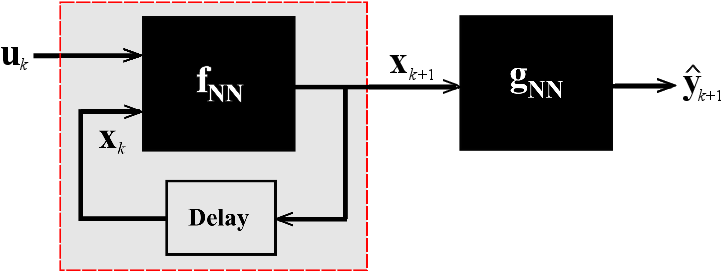}
 		\caption{{   SSNNO block diagram.}}
 		\label{figSSNNO}		
 	\end{center}
 \end{figure}

The proposed   
 SSNNO shares the same structural form as the SSNN \cite{bJZ98}; the key difference lies in its training,  which simultaneously identifies parameters of the network and the reduced model order. 
Given training data consisting of a sufficiently rich (informative)  
input and output sequences of size ${N}$:
\begin{equation}
      \label{equy}  
      \begin{aligned}
     \mathbf{U}&=\left[\begin{matrix} \mathbf{u}_{0} & \mathbf{u}_{1} & \dots & \mathbf{u}_{{N}-1}\end{matrix}\right],\hspace{0.5cm} \mathbf{U}\in \mathbb{R}^{m \times N}\\\mathbf{Y}&=\left[\begin{matrix} \mathbf{y}_{0} & \mathbf{y}_{1} & \dots & \mathbf{y}_{{N}-1}\end{matrix}\right], \hspace{0.5cm} \mathbf{Y}\in \mathbb{R}^{p \times N}
     \end{aligned}
    \end{equation}
the SSNNO attempts to learn a state-space model:
\begin{equation}
  \label{eqssnno}   
  \begin{aligned} {\mathbf{x}}_{k+1}&=\mathbf{f}_{\text{NN}}({\mathbf{x}}_{k},\mathbf{u}_{k};\bm{\theta}_{\mathbf{f}})=
\mathbf{f}_{\text{NN}}({\mathbf{x}}_{k},\mathbf{u}_{k}) \\
\widehat{\mathbf{y}}_{k}&=\textbf{g}_{\text{NN}}({\mathbf{x}}_{k};\bm{\theta}_{\textbf{g}})=\textbf{g}_{\text{NN}}({\mathbf{x}}_{k})
 \end{aligned}
\end{equation}
where ${\mathbf{x}}_{k}\in \mathbb{R}^{d}$ is the state vector predicted using SSNNO, $\widehat{\mathbf{y}}_{k} \in \mathbb{R}^{p}$ is the predicted output, 
$\mathbf{f}_{\text{NN}}:\mathbb{R}^{d} \times \mathbb{R}^{m} \rightarrow \mathbb{R}^{d},$ $\textbf{g}_{\text{NN}}:\mathbb{R}^{d} \rightarrow \mathbb{R}^{p}$ are the state and output functions, each modeled as a (possibly deep) NN, with $\bm{\theta}_{\mathbf{f}} \in \mathbb{R}^{n_{f}},$ $\bm{\theta}_{\textbf{g}} \in \mathbb{R}^{n_{g}}$ representing the corresponding weights and biases. A block diagram of SSNNO (or SSNN) is given in Fig. \ref{figSSNNO}.

The state and output functions are represented with subnetworks: %evaluated as a composition of layers:
\begin{equation}
 \label{eqfnngnn}
    \begin{aligned}
        \mathbf{f}_{\text{NN}}&=\mathbf{f}_{{L}}({{\dots}}\mathbf{f}_{2}(\mathbf{f}_{1}({\mathbf{x}}_{k},\mathbf{u}_{k})){{\dots}})\\\textbf{g}_\text{NN}&=\textbf{g}_{{H}}({{\dots}}\textbf{g}_{2}(\textbf{g}_{1}({\mathbf{x}}_{k})){{\dots}})
    \end{aligned}
\end{equation}
where  $\mathbf{f}_{i}$ and $\textbf{g}_{i}$ are layers of state and output  subnetworks  represented as:
\begin{equation}
 \label{eqfi}
 \begin{aligned}
 \mathbf{f}_{i}(.)&=\bm{\bm{\sigma}}_{\mathbf{f}_i}(\mathbf{A}_{\mathbf{f}_i}(.)+\mathbf{b}_{\mathbf{f}_i})\\
 \textbf{g}_{i}(.)&=\bm{\bm{\sigma}}_{\textbf{g}_i}(\mathbf{A}_{\textbf{g}_i}(.)+\mathbf{b}_{\textbf{g}_i})
  \end{aligned}
\end{equation}
$\bm{\bm{\sigma}}_{\mathbf{f}_i}:\mathbb{R}^{{l_{i}}}\rightarrow \mathbb{R}^{l_i},$ $\bm{\sigma}_{\textbf{g}_i}:\mathbb{R}^{{h_{i}}}\rightarrow \mathbb{R}^{h_{i}}$ contain activation functions for the  $i^{th}$ layer of  state and output subnetworks, with corresponding weight and bias parameters being $\mathbf{A}_{\mathbf{f}_i} \in \mathbb{R}^{l_{i} \times l_{{i-1}} },$ $\mathbf{b}_{\mathbf{f}_i} \in \mathbb{R}^{l_{i}}$, $\mathbf{A}_{\textbf{g}_i} \in \mathbb{R}^{h_{i} \times h_{i-1} }$, $\mathbf{b}_{\textbf{g}_i} \in \mathbb{R}^{h_{i}},$ respectively.  
Let the state and output sequences predicted using SSNNO be defined as 
\begin{equation}
 \label{eqxnny}
 \begin{aligned}
{\mathbf{X}}&=\Big[ {\mathbf{x}}_{0} \hspace{.3cm} \mathbf{f}_{\text{NN}}({\mathbf{x}}_{0},\mathbf{u}_{0}) \hspace{.3cm} \mathbf{f}_{\text{NN}}\big(\mathbf{f}_{\text{NN}}({\mathbf{x}}_{0},\mathbf{u}_{0}),\mathbf{u}_{1}\big) \hspace{.3cm} \dots \hspace{.3cm}\mathbf{f}_{\text{NN}}\big(\mathbf{f}_{\text{NN}}\big(\dots\mathbf{f}_{\text{NN}}({\mathbf{x}}_{0},\mathbf{u}_{0})\dots \big), \mathbf{u}_{{N}-2} \big) \Big]\\
\widehat{\mathbf{Y}}&=\left[\begin{matrix} \textbf{g}_{\text{NN}}({\mathbf{x}}_{0}) & \textbf{g}_{\text{NN}}({\mathbf{x}}_{1}) & \dots & \textbf{g}_{\text{NN}}({\mathbf{x}}_{{N}-1})\end{matrix}\right].
\end{aligned}
\end{equation}
which are obtained by simulating the SSNNO for the training data inputs $\mathbf{U}$ in Eq. (\ref{equy}) for some initial state ${\mathbf{x}}_{0}$.
For the predicted state vector sequence $\mathbf{X}$, denote the sample mean vector and sample covariance matrix  as $\bar{\mathbf{x}}$ and $\mathbf{V}_{{\mathbf{x}}}$, respectively.
{The diagonal elements of $\mathbf{V}_{\mathbf{x}}$ correspond to sample variances of the predicted state variables: 
\begin{equation}
\label{eqvxoi}
V_{{{x}}_i}=\mathbf{V}_{{\mathbf{x}}}(i,i), \hspace{1cm} i=1,2,\dots,d.
\end{equation}}
%We now define  SSNNO as follows:\\
\begin{definition}
\label{defssnno}
    An SSNNO is a state-space model of the form as in Eq. \eqref{eqssnno} for which the sample variances of the predicted states Eq. (\ref{eqvxoi}) computed for training data Eq. (\ref{equy}), satisfy:
    \begin{equation} \label{ordering}
        V_{x_{1}}\geq V_{x_{2}}\geq \dots \geq V_{x_{d}} \geq 0.
    \end{equation}
\end{definition}
 Note that both SSNN and SSNNO are represented as in Eq. (\ref{eqssnno}). However, in SSNNO Eq. \eqref{ordering} is explicitly enforced. We propose to achieve variance-ordering Eq. \eqref{ordering} by incorporating a variance regularization term in the loss function, as discussed next. Incorporation of variance regularization not only ensures variance-ordering but also suppresses the variances of the trailing states to negligible levels, without significantly increasing the squared prediction error (SPE). 
 \subsection{SSNNO Training: Variance-based State Regularization}
 Define ${\bm{\theta}} \in \mathbb{R}^{{n_{f}}+{n_{g}}+d}$ which contains the weight parameters for $\mathbf{f}_{\text{NN}},$ $\textbf{g}_{\text{NN}}$ as well as the initial state ${\mathbf{x}}_{0} \in \mathbb{R}^{d} $ as: 
\begin{equation}
    \label{eqpar}
    \bm{\theta}= {{ \left[\begin{matrix} \bm{\theta}_{\mathbf{f}}^{\top} & ~\bm{\theta}_{\textbf{g}}^{\top} & ~{\mathbf{x}}_{0}^{\top} \end{matrix}\right]}}^{\top}.
\end{equation}
The loss function for SSNNO is constructed by seeking to minimize weighted sample variance and parameter regularization terms, in addition to the SPE, as follows:
{\begin{equation}
 \label{eqloss}
    J = \underbrace{{\parallel \mathbf{Y}-\widehat{\mathbf{Y}} \parallel }_{F}^{2}}_{J_{y}}+\alpha \underbrace{  {\parallel \mathbf{W}^{\frac{1}{2}}[\mathbf{X}-\bar{\mathbf{X}}] \parallel }_{F}^{2}}_{J_{v}}  +\beta \underbrace{ {\parallel \bm{\theta}_{\textbf{g}}  \parallel }_{2}^{2}}_{J_{g}}  
\end{equation}}
\noindent where $\alpha,\beta>0$ are hyperparameters. $J_{y}$ denotes the SPE which is the loss function of conventional SSNNs\cite{bJS95,bJZ98,bKK18}, while $J_{g}$ is the parameter-regularization term  \cite{bYT22} associated with the output subnetwork.  $J_{v}$ regularizes the sample variances of the SSNNO states in the state subnetwork Eq. \eqref{eqssnno} and involves a diagonal {weighting matrix} whose non-negative elements are arranged in increasing order:
\begin{equation}
\label{weight_matrix}
    \mathbf{W}= diag(w_{1},\dots,w_{l}), \hspace{0.5cm} 0\leq w_{1} <\dots < w_{d}< \infty.
\end{equation}

%Thus, the trained SSNNO ensures that the states are ordered by their variances: the first of the $d$ states exhibits the highest variance for the training data, and each subsequent state having a progressively smaller variance. 
To simplify the choice of weights, the elements of $\mathbf{W}$ are specifically chosen as fixed constants in this paper, $\mathbf{W}=diag(1,2,\dots,d)$. As a result, $\alpha$ and $\beta$ are the only hyperparameters in the loss function.  Note that $J_v$  in SSNNO  tends to promote states with negligible variances during training, reflecting the trade-off between the SPE and the model order.  
In the proposed SSNNO, the parameters of the state subnetwork are excluded from the parameter regularization term. This is motivated by the fact that applying both state variance regularization and state subnetwork parameter regularization may lead to excessively small state values, which in turn would be counterbalanced by larger parameters in the output subnetwork. To mitigate this issue, only the parameters of the output subnetwork are included in the parameter regularization term. An alternative approach would be to apply parameter regularization to both the state and output subnetworks, with distinct scaling factors for each. However, this would increase the number of hyperparameters in the loss function. 
The training problem for SSNNO is stated as:
\begin{problem}[SSNNO Training]
     \label{SSNNO_problem}
    Given training data in Eq. \eqref{equy}, estimate parameters $\bm{\theta}$ in Eq. \eqref{eqpar}, thereby obtaining the trained SSNNO in Eq. \eqref{eqssnno}, by minimizing the loss function in Eqs. (\ref{eqloss}), (\ref{weight_matrix}) with  suitable choices of $\alpha,\beta$,  as follows:
\begin{equation}
\label{eqopto}
\bm{\theta}^{*}=\arg\underset{\bm{\theta}}{\min} \hspace{.2cm} J.
\end{equation}    
\end{problem}
% \begin{remark}
% %The point of departure between SSNN and SSNNO is the inclusion of objective $J_{v}$ in Eq. (\ref{eqloss}). 
% The trade-off between $J_{y}$ and $J_{v}$ in Eq. (\ref{eqloss}) ensures that only those states that play a significant role in minimizing the SPE are allowed significant sample variances. Thus, any overparameterization of the number of states owing to the user-specified order $d$ is identified by the criterion for determination of the reduced model order $s$, as will be discussed in Section \ref{model_order}.
% \end{remark} 
%Elements of $\bm{\theta}^{*}$ specify parameters of trained subnetworks $\mathbf{f}_{\text{NN}}$ and $\textbf{g}_{\text{NN}}$ which, in turn, yield the state-space model with the user-specified order $d$ as in Eq. (\ref{eqssnno}).
\begin{remark}
    Definition \ref{defssnno} requires that the variance of the states of the SSNNO be ordered. The SSNNO training in Problem \ref{SSNNO_problem} is formulated not only to enforce this ordering but also to promote states with negligible variances. While Theorem \ref{lemmassnno} (presented in the next section) guarantees that the global minimizer for Problem \ref{SSNNO_problem}  yields an SSNNO in the sense of Definition \ref{defssnno}, standard NN training algorithms often converge only to a local minimum of the nonconvex objective Eq. (\ref{eqloss}) and the resulting trained network may not satisfy the variance-ordering property required in Definition \ref{defssnno}. For such cases, a practical method to transform any locally optimized solution of Eq. (\ref{eqopto}) into a variance-ordered SSNNO is described in Section \ref{sec_practicalmethod}.
\end{remark}
%The optimization problem for SSNNO in Eq. (\ref{eqopto}) is inherently nonconvex. Theorem \ref{lemmassnno} (presented in the next section) establishes that the global minimizer of Eq. (\ref{eqopto}) enforces an ordering of sample variances of the states. Although several recent works have proposed effective numerical methods for handling nonconvex optimization problems \cite{bFM25,bAB25,bKA25},  conventional NN training techniques do not guarantee convergence to the global solution. To circumvent this issue, Section \ref{sec_practicalmethod} introduces a practical method for obtaining an SSNNO with guaranteed ordering of the state variances from a network that converges only to a local minimum in Eq. (\ref{eqopto}). This aspect is elaborated in the next section. 
We now discuss a procedure to obtain a reduced-order model from the trained SSNNO.

\subsection{Reduced Order SSNNO (R-SSNNO)} \label{model_order}
 Assume that a trained SSNNO, as discussed in the previous section, is available. Thus, the predicted state variables {$x_{1},x_{2},\dots,x_{d}$} are ordered in terms of decreasing variance: $V_{x_{1}}\geq V_{x_{2}}\geq \dots \geq V_{x_{d}} \geq 0$. 
  To determine the reduced model order $s\leq d$,  we find the number of state variables in the trained SSNNO that exhibit significant variance. This is achieved by categorizing the states of the SSNNO into significant and residual states as follows:
\begin{definition}
    For a given tolerance $\delta \geq 0$ and a trained SSNNO obtained by solving Problem \ref{SSNNO_problem}, the state variable $x_{i},$ $i=1,2,\dots,d$  is labeled a {significant state}, if 
    \begin{equation}
\label{eqvar}
V_{{x_{}}_i}>\delta
\end{equation}
    otherwise, it is labeled a {residual state}. Thus, the states of the trained SSNNO are partitioned as: $\mathbf{x} =  [\underbrace{x_{1} \cdots x_{s}}_{V_{x_{i}} > \delta} | \underbrace{x_{{s+1}} \cdots x_{d}}_{V_{x_{i}} \leq \delta} ]^{\top} = \left[\begin{matrix} \mathbf{x}_{s} \\ \mathbf{x}_{r} \end{matrix}\right]$
   where $\mathbf{x}_{s} \in \mathbb{R}^{s}, \mathbf{x}_{r} \in \mathbb{R}^{d-s},$ correspond to the significant and residual states. %Moreover, the sample mean and covariance for the significant states are $\bar{\mathbf{x}}_\mathbf{s},\mathbf{V}_{\mathbf{x}_{s}}$ and for the residual states are $\bar{\mathbf{x}}_\mathbf{r},\mathbf{V}_{\mathbf{x}_{r}}$.
\end{definition}
Model order reduction is achieved by retaining only the $s$ significant states % with dominant sample variances (discussed in Section \ref{model_order})
and absorbing the remaining $d-s$ residual states into the SSNNO model parameters
yielding an R-SSNNO model with order $s \leq d$ from the trained SSNNO without requiring any retraining.
 This is achieved by suitably partitioning the parameters of the first and last layers of subnetwork $\mathbf{f}_{\text{NN}}$ of the trained SSNNO, namely, $\mathbf{A}_{\mathbf{f}_{1}},$ $\mathbf{A}_{\mathbf{f}_{{L}}},$ based on the reduced model order $s$:
\begin{equation}
\label{eqgroup} 
\mathbf{A}_{\mathbf{f}_1} ={{\left[\begin{matrix} \mathbf{A}_{\mathbf{f}_{1_{s}}} &  \mathbf{A}_{\mathbf{f}_{1_{r}}}   &\mathbf{A}_{\mathbf{f}_{1_{u}}}  \end{matrix}\right]}},
   \mathbf{A}_{\mathbf{f}_{{L}}} ={{\left[\begin{matrix} \mathbf{A}_{\mathbf{f}_{{L}{s}}} \\   \mathbf{A}_{\mathbf{f}_{{L}{r}}} \end{matrix}\right]}},\mathbf{b}_{\mathbf{f}_{{L}}} ={{\left[\begin{matrix} \mathbf{b}_{\mathbf{f}_{{L}{s}}} \\   \mathbf{b}_{\mathbf{f}_{{L}{r}}} \end{matrix}\right]}}
\end{equation}
where 
$\mathbf{A}_{\mathbf{f}_{1_s}} \in \mathbb{R}^{l_{1} \times s},$
$\mathbf{A}_{\mathbf{f}_{1_r}} \in \mathbb{R}^{l_{1} \times (d-s)},$
$\mathbf{A}_{\mathbf{f}_{1_u}} \in \mathbb{R}^{l_{1} \times m},$
$\mathbf{A}_{\mathbf{f}_{{L}_s}} \in \mathbb{R}^{ s \times l_{{L-1}}},$
$\mathbf{A}_{\mathbf{f}_{{L}_r}} \in \mathbb{R}^{ (d-s) \times l_{{L-1}}}$, $\mathbf{b}_{\mathbf{f}_{{L}_s}} \in \mathbb{R}^{s}$, $\mathbf{b}_{\mathbf{f}_{{L}_r}} \in \mathbb{R}^{d-s}$. Similarly, partition parameters of the first layer of $\textbf{g}_{\text{NN}}$ as follows: $\mathbf{A}_{\textbf{g}_{1}} =\left[\begin{matrix} \mathbf{A}_{\textbf{g}_{1_{s}}} &   \mathbf{A}_{\textbf{g}_{1_{r}}} \end{matrix}\right]$
where $\mathbf{A}_{\textbf{g}_{1_s}} \in \mathbb{R}^{ h_{1} \times s },\mathbf{A}_{\textbf{g}_{1_r}} \in \mathbb{R}^{ h_{1} \times (d-s) }.$ 
Using this partitioning and Eq. (\ref{eqfnngnn}), the state and output equations in Eq. (\ref{eqssnno}) are rewritten as:
\begin{equation}
\label{eqsysder} 
\begin{aligned}
\mathbf{x}_{{k+1}}=&\left[\begin{matrix} \mathbf{x}_{s_{k+1}} \\ \mathbf{x}_{r_{k+1}} \end{matrix}\right] =
 \mathbf{f}_{{L}}({{\dots}}\mathbf{f}_{2}(\bm{\bm{\sigma}}_{\mathbf{f}_1}(\mathbf{A}_{\mathbf{f}_{1_s}}\mathbf{x}_{s_k}+\mathbf{A}_{\mathbf{f}_{1_r}}\mathbf{x}_{r_k} + \mathbf{A}_{\mathbf{f}_{1_u}} \mathbf{u}_{k} + \mathbf{b}_{\mathbf{f}_1} )){{\dots}})\\
\widehat{\mathbf{y}}_{k} = 
 &~  \textbf{g}_{H}({{\dots}}\textbf{g}_{2}(\bm{\bm{\sigma}}_{\textbf{g}_1}(\mathbf{A}_{\textbf{g}_{1_s}}\mathbf{x}_{s_k}+\mathbf{A}_{\textbf{g}_{1_r}}\mathbf{x}_{r_k} + \mathbf{b}_{\textbf{g}_1} )){{\dots}})
\end{aligned}
\end{equation}
Since the sample variance of residual states is small, that is, ${V}_{{x}_i} \le \delta, $ for $i=s+1,\cdots,d$, these states do not exhibit significant variation over the training data, and may be approximated by their mean $\bar{\mathbf{x}}_\mathbf{r}$. Thus,
\begin{equation} \label{mor_approx}
    \mathbf{x}_{r_{k+1}} \approx  \mathbf{x}_{r_k} \approx \bar{\mathbf{x}}_{r}.
\end{equation}
The above approximation simplifies Eq. (\ref{eqsysder}) to yield the R-SSNNO model of order $s$:
\begin{equation}
\label{eqsysid} 
\begin{aligned}
\mathbf{x}_{s_{k+1}}   & \approx
  \mathbf{f}_{{{L}_s}}({{\dots}}
\mathbf{f}_{2}(\bm{\bm{\sigma}}_{\mathbf{f}_1}(\mathbf{A}_{\mathbf{f}_{1_s}}\mathbf{x}_{s_k}  + \mathbf{A}_{\mathbf{f}_{1_u}} \mathbf{u}_{k} + \mathbf{b}_{\mathbf{f}_{1_s}} )){{\dots}})\\& =\mathbf{f}_{s}(\mathbf{x}_{s_k},\mathbf{u}_{k})\\
\widehat{\mathbf{y}}_{s_k} & =  \textbf{g}_{H}({{\dots}}\textbf{g}_{2}(\bm{\bm{\sigma}}_{\textbf{g}_1}(\mathbf{A}_{\textbf{g}_{1_s}}\mathbf{x}_{s_k}+ \mathbf{b}_{\textbf{g}_{1_s}} )){{\dots}})\\&=\textbf{g}_{s}(\mathbf{x}_{s_k})
\end{aligned}
\end{equation}
where $\mathbf{f}_{{L}_s}(.)= \bm{\bm{\sigma}}_{\mathbf{f}_{{L}_s}}(\mathbf{A}_{\mathbf{f}_{{L}_{s}}}(.)+\mathbf{b}_{\mathbf{f}_{{L}_{s}}})$, $\bm{\bm{\sigma}}_{\mathbf{f}_{{L}_s}}$ corresponds to the activation functions of the first $s$ nodes in the ${L}^{th}$ layer,
$\widehat{\mathbf{y}}_{s_k} \in \mathbb{R}^{p}$ is the output predicted using the R-SSNNO model, $\mathbf{b}_{\mathbf{f}_{1_s}}=\mathbf{b}_{\mathbf{f}_1} +  \mathbf{A}_{\mathbf{f}_{1_r}}\bar{\mathbf{x}}_{r},$ and $\mathbf{b}_{\textbf{g}_{1_s}}=\mathbf{b}_{\textbf{g}_1} +  \mathbf{A}_{\textbf{g}_{1_r}}\bar{\mathbf{x}}_{r}$. Note that the SSNNO model is approximated using Eq. (\ref{mor_approx}), which results in an R-SSNNO model with order $s\leq d$.  The reduced model order can be controlled by adjusting $\delta$ in Eq. (\ref{eqvar}). 
\section{Theoretical results on SSNNO}
{One of the key features of the proposed SSNNO obtained by solving Eq. (\ref{eqopto}) is that the states are sorted in decreasing order of variance. 
In this section, we establish the existence of such an SSNNO. 
We also show that this variance-ordering can be achieved while ensuring that the corresponding increase in SPE of SSNNO relative to SSNN can be upper-bounded by a user-specified tolerance. 
To this end, a supporting lemma that establishes equivalence between SSNN as in Eq. (\ref{eqssnno}) and a permuted SSNN (P-SSNN) based on permutation matrix $\mathbf{T}$ as in Fig. \ref{figPSSNN}, wherein the states of the SSNN are arbitrarily permuted, is presented next:
\begin{figure}[h!]
	
	\begin{center}
		\includegraphics [scale=0.75] {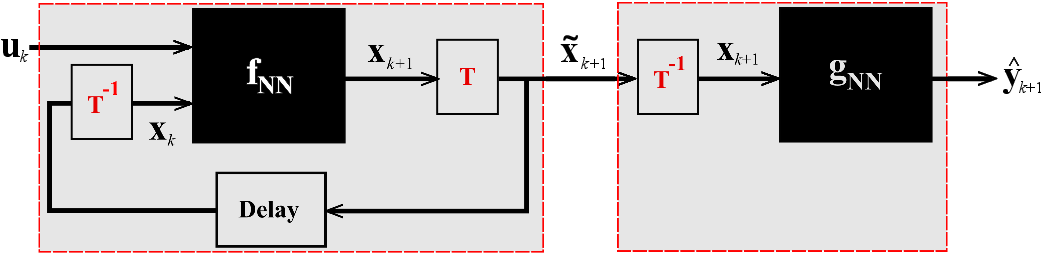}
		\caption{{ P-SSNN from  SSNN.}}
	\label{figPSSNN}	
	\end{center}
\end{figure}
\begin{lemma}
   \label{lemmaperm}
    Consider an SSNN as in Eq. (\ref{eqssnno}), with known states ${\mathbf{x}}_{k}$ and  parameters $\bm{\theta}$. Then there exists an equivalent SSNN (labeled as P-SSNN) with permuted states  $\widetilde{\mathbf{x}}_k = \mathbf{T}{\mathbf{x}}_{k}$  and parameters $\widetilde{\bm{\theta}}$ which are also some permutation of $\bm{\theta}$,  corresponding to an arbitrary permutation matrix $\mathbf{T}$, such that P-SSNN and SSNN have identical input-output mapping. 
\end{lemma}
\begin{proof}
    Consider an SSNN with subnetworks $\mathbf{f}_{\text{NN}}, \textbf{g}_{\text{NN}}$ with corresponding parameters $\bm{\theta}_\mathbf{f}, \bm{\theta}_\textbf{g}$. For a known state ${\mathbf{x}}_{k}$, input $\mathbf{u}_{k}$, the SSNN prediction of state  ${\mathbf{x}}_{k+1}$ and output $\widehat{\mathbf{y}}_{k+1}$ are given by Eq. (\ref{eqssnno}):
     \begin{align}
     {\mathbf{x}}_{k+1} &= \mathbf{f}_{\text{NN}}({\mathbf{x}}_{k},\mathbf{u}_{k}) = \mathbf{f}_{{L}}(\cdots \mathbf{f}_{2}(\mathbf{f}_{1}({\mathbf{x}}_{k},\mathbf{u}_{k}))\cdots) \nonumber\\
 &=\bm{\sigma}_{\mathbf{f}_{{L}}}\Bigg(\mathbf{A}_{\mathbf{f}_{{L}}}\bm{\sigma}_{\mathbf{f}_{{L}-1}}\bigg(\cdots \bm{\sigma}_{\mathbf{f}_{1}}\bigg(\mathbf{A}_{\mathbf{f}_{1}}\begin{bmatrix}
        {\mathbf{x}}_{k} \\ \mathbf{u}_{k}
     \end{bmatrix}
    +\mathbf{b}_{\mathbf{f}_{1}}\bigg)   \cdots + 
    \mathbf{b}_{\mathbf{f}_{{L}-1}} \bigg) + \mathbf{b}_{\mathbf{f}_{{L}}} \Biggl ) \label{eq_ssnnstate}\\
    \widehat{\mathbf{y}}_{k+1} &= \textbf{g}_{\text{NN}}({\mathbf{x}}_{k+1})=\textbf{g}_{H}(\cdots \textbf{g}_{2}(\textbf{g}_{1}({\mathbf{x}}_{k+1}))\cdots) \nonumber\\
 &=\bm{\sigma}_{\textbf{g}_{H}}\bigg(\mathbf{A}_{\textbf{g}_{H}}\bm{\sigma}_{\textbf{g}_{{H}-1}}\big(\cdots \bm{\sigma}_{\textbf{g}_{1}}(\mathbf{A}_{\textbf{g}_{1}}{\mathbf{x}}_{k+1}
    +\mathbf{b}_{\textbf{g}_{1}})  \cdots + 
    \mathbf{b}_{\textbf{g}_{{H}-1}} \big) + \mathbf{b}_{\textbf{g}_{H}} \bigg). \label{eqt3}
 \end{align}  
Using the permutation matrix $\mathbf{T}$, the state and observation functions of the SSNN can be rewritten in terms of permuted states 
$\widetilde{\mathbf{x}}_k = \mathbf{T}{\mathbf{x}}_{k}$ as:
\begin{align}
    \widetilde{\mathbf{x}}_{k+1} &= \mathbf{T}\mathbf{f}_{\text{NN}}(\mathbf{T}^{-1}\widetilde{\mathbf{x}}_k,\mathbf{u}_{k}) \nonumber   \\
&=\mathbf{T}\bm{\sigma}_{\mathbf{f}_{{L}}}\Bigg(\mathbf{A}_{\mathbf{f}_{{L}}}\bm{\sigma}_{\mathbf{f}_{{L}-1}}\bigg(\cdots \bm{\sigma}_{\mathbf{f}_{1}}\big(\mathbf{A}_{\mathbf{f}_{1}}\begin{bmatrix}
    \mathbf{T}^{-1} & \mathbf{I}_{m}
\end{bmatrix}\begin{bmatrix}
       \widetilde{\mathbf{x}}_k \\ \mathbf{u}_{k}
    \end{bmatrix}  +\mathbf{b}_{\mathbf{f}_{1}} \big) \cdots + 
   \mathbf{b}_{\mathbf{f}_{{L}-1}} \bigg) + \mathbf{b}_{\mathbf{f}_{{L}}} \Bigg)  \label{eqt1}
\end{align}
\begin{align}   
   \widehat{\mathbf{y}}_{k+1} &= \textbf{g}_{\text{NN}}(\widehat{\mathbf{x}}_{k+1})=\textbf{g}_{\text{NN}}(\mathbf{T}^{-1}\widetilde{\mathbf{x}}_{k+1})\\
& = \bm{\sigma}_{\textbf{g}_{H}}\bigg(\mathbf{A}_{\textbf{g}_{H}}\bm{\sigma}_{\textbf{g}_{{H}-1}}\big(\cdots \bm{\sigma}_{\textbf{g}_{1}}(\mathbf{A}_{\textbf{g}_{1}}\mathbf{T}^{-1}\widetilde{\mathbf{x}}_{k+1}
   +\mathbf{b}_{\textbf{g}_{1}}) \cdots + \mathbf{b}_{\textbf{g}_{{H}-1}}\big) + \mathbf{b}_{\textbf{g}_{H}}\bigg). \label{eqt2}
\end{align}
%The output $\widehat{\mathbf{y}}_{k+1}$ is invariant under the permutation $\mathbf{T},$ since $\mathbf{T}$ is invertible and  Eq. (\ref{eqt1}) constitutes a coordinate transformation (see \cite{Nijmeijer2016} Eq. (5.11)). 
Define
\begin{align}  \widetilde{\bm{\bm{\sigma}}}_{\mathbf{f}_{{L}}}(\cdot) &= \mathbf{T}\bm{\sigma}_{\mathbf{f}_{{L}}}(\cdot), \hspace{0.7cm} 
    \widetilde{\mathbf{A}}_{\mathbf{f}_{1}} = \mathbf{A}_{\mathbf{f}_{1}}\begin{bmatrix}
    \mathbf{T}^{-1} & \mathbf{I}_{m}
\end{bmatrix}  \label{eq_af1} \\
\widetilde{\mathbf{A}}_{\textbf{g}_{1}} &=\mathbf{A}_{\textbf{g}_{1}}\mathbf{T}^{-1}. \label{eq_ag1}
\end{align}
Construct P-SSNN as an SSNN with subnetworks $\widetilde{\mathbf{f}}_{\text{NN}},\widetilde{\textbf{g}}_{\text{NN}}$ defined as:
\begin{align}
\widetilde{\mathbf{f}}_i & = \left\{
\begin{array}{ll}
\bm{\sigma}_{\mathbf{f}_{1}} (\widetilde{\mathbf{A}}_{\mathbf{f}_{1}}(.) + \mathbf{b}_{\mathbf{f}_{1}}),~&~i=1 \\
\bm{\bm{\sigma}}_{\mathbf{f}_i} ( \mathbf{A}_{\mathbf{f}_i}(.) + \mathbf{b}_{\mathbf{f}_i}),~&~i=2,3,...,{L}-1 \\
\widetilde{\bm{\bm{\sigma}}}_{\mathbf{f}_{{L}}} ( \mathbf{A}_{\mathbf{f}_{{L}}}(.) + \mathbf{b}_{\mathbf{f}_{{L}}}),~&~i={L}
\end{array}  \right.  \label{eq_ftildei} \\
\widetilde{\textbf{g}}_i & = \left\{
\begin{array}{ll}
\bm{\bm{\sigma}}_{\textbf{g}_{1}} (\widetilde{\mathbf{A}}_{\textbf{g}_{1}}(.) + \mathbf{b}_{\textbf{g}_{1}}),~&~i=1 \\
\bm{\sigma}_{\textbf{g}_i} ( \mathbf{A}_{\textbf{g}_i}(.) + \mathbf{b}_{\textbf{g}_i}),~&~i=2,3,...,{H}. 
\end{array} \right\}  \label{eq_gtildei}
\end{align}
Thus, P-SSNN has the same set of parameters as SSNN, other than in the first and last layers of the state subnetwork and the first layer of the output subnetwork. For these layers, permuted activation functions/ parameters of SSNN  are used. Eqs. (\ref{eqt1}),(\ref{eqt2}) can then be rewritten as:
\begin{align}
\widetilde{\mathbf{x}}_{k+1} & =\widetilde{\mathbf{f}}_{{L}}({{\dots}}\widetilde{\mathbf{f}}_{2}(\widetilde{\mathbf{f}}_{1}(\widetilde{\mathbf{x}}_{k},\mathbf{u}_{k}))  = \widetilde{\mathbf{f}}_{\text{NN}}(\widetilde{\mathbf{x}}_{k},\mathbf{u}_{k}) \label{eq_fnntilde} \\
\widehat{\mathbf{y}}_{k+1} &=\widetilde{\textbf{g}}_{{H}}({{\dots}}\widetilde{\textbf{g}}_{2}(\widetilde{\textbf{g}}_{1}(\widetilde{\mathbf{x}}_{k}))  = \widetilde{\textbf{g}}_{\text{NN}}(\widetilde{\mathbf{x}}_{k+1}). \label{eqt4}
\end{align}
It is seen from Eqs. (\ref{eqssnno}),(\ref{eqt1}),  (\ref{eqt2}),  that for a state ${\mathbf{x}}_{k}$ and input $\mathbf{u}_{k}$, the output $\widehat{\mathbf{y}}_{k+1}$ from SSNN is identical to the output obtained from P-SSNN with  state $\widetilde{\mathbf{x}}_k=\mathbf{T}{\mathbf{x}}_{k}$ and input $\mathbf{u}_{k}$. 
\end{proof}}
{The following lemma from literature is needed to show the existence of SSNNO with ordered variance of the state vector:
\begin{lemma}[Rearrangement inequality (Theorem 368 in \cite{bGH34})]
\label{lemmarearr}
    For every pair of $n$--tuple of reals:
    \begin{equation}
        a_{1}\leq a_{2}\leq \dots \leq a_{n}, \hspace{1cm} b_{1} \leq b_{2} \leq \dots \leq b_{n}
    \end{equation}
    and every set of numbers $\{q_{1},q_{2},\dots,q_{n}\}$ obtained by permuting the elements of the set $\{1,2,\dots,n\}$, the following inequality holds:
    \begin{equation}
    \label{eqrearr}
    \begin{aligned}
        &a_{1}b_{n}+a_{2}b_{n-1}+\dots +a_{n}b_{1} \leq a_{1} b_{q_1} +a_{2} b_{q_2}+\dots + a_{n} b_{q_n} \\&\leq a_{1}b_{1}+a_{2}b_{2}+\dots a_{n}b_{n}.
            \end{aligned}
    \end{equation}
\end{lemma}
% \begin{proof}
%  Refer to \cite{bGH34}.    
% \end{proof}
Next, using Lemmas \ref{lemmaperm} and \ref{lemmarearr}, a guarantee on the existence of SSNNO with variance-ordered states is provided in Theorem \ref{lemmassnno}. The proof relies on the availability of the global minimizer of Eq. (\ref{eqopto}). A practical solution for obtaining an SSNNO with variance-ordered states using a local minimizer is presented later. 
\begin{theorem}[Existence of SSNNO]
 \label{lemmassnno}
For every $\mathbf{W}$ with $0\leq w_{1} < w_{2} < \dots <w_{d}$, the global minimizer to Eq. (\ref{eqopto}) satisfies:
\begin{equation}
\label{eqvarord}
    \begin{aligned}
       V_{x_1} \geq V_{x_2} \geq \dots \geq V_{x_d} \geq 0.
    \end{aligned}
\end{equation}
\end{theorem}
\begin{proof}
The proof is by contradiction. Let the global minimizer of Eq. (\ref{eqopto}) result in an SSNNO with non-ordered variance of states $\widehat{\mathbf{x}}$. Let the corresponding optimal loss function (Eqs. (\ref{eqloss}),(\ref{eqopto})) be:
\begin{align}
    \widehat{J} & = \widehat{J}_y + \alpha \widehat{J}_v + \beta \widehat{J}_g. 
\label{eqn:lossSSNNO}
\end{align}
Sort the states of this SSNNO in decreasing order of variance and store the sorting index in vector $\mathbf{z}$. Define a permutation matrix:
\begin{align}
\label{eqpermat}
\mathbf{T} & = \mathbf{I}_d(\mathbf{z},:)
\end{align}
where $\mathbf{I}_d(\mathbf{z},:)$ is obtained by rearranging rows of the identity matrix $\mathbf{I}_d$ according to the index vector $\mathbf{z}$. 
With the permutation matrix $\mathbf{T}$, define an equivalent P-SSNNO as in Lemma 1, with permuted states:
\begin{equation}
\label{eqstateperm}
\widetilde{\mathbf{x}}_{k}=\mathbf{T}{\mathbf{x}}_{k}.
\end{equation}
Let the loss function for the P-SSNNO be
\begin{align}
    \widetilde{J} & = \widetilde{J}_y + \alpha \widetilde{J}_v + \beta \widetilde{J}_g. 
\label{eqn:lossP-SSNNO}
\end{align}
From Lemma \ref{lemmaperm}, it then follows that
\begin{equation}
\label{eqn:identicalSPE}
\widetilde{J}_y = \widehat{J}_y, \hspace{0.5cm}
\widetilde{J}_g = \widehat{J}_g.
\end{equation}
Eq. (\ref{eqn:identicalSPE}) follows since P-SSNNO and SSNNO have identical outputs, and parameters of P-SSNNO are permuted parameters of SSNNO. 
The sample variance regularization term of P-SSNNO is:
\begin{align}   
\widetilde{J}_{v}&= \operatorname{\operatorname{trace}}([\widetilde{\mathbf{X}}-\bar{\widetilde{\mathbf{X}}}]^{\top}\mathbf{W}[\widetilde{\mathbf{X}}-\bar{\widetilde{\mathbf{X}}}]) \nonumber\\ &= ({N}-1)(w_{1}V_{x_{z_1}}+w_{2}V_{x_{z_2}}+ \dots+w_{d}V_{x_{z_d}}) \nonumber\\
& \leq ({N}-1)(w_{1}V_{x_1}+
\dots+w_{d}V_{x_d}) =  \widehat{J}_v
\label{eqn:rearrangementuse}
%\\
%\implies \widetilde{J}_v & \leq \widehat{J}_v
%\label{eqn:lowervariance}
\end{align}
where inequality (\ref{eqn:rearrangementuse}) follows from rearrangement inequality (Lemma 2). Thus, from Eqs. (\ref{eqn:lossSSNNO}), (\ref{eqn:lossP-SSNNO}-\ref{eqn:identicalSPE}) and (\ref{eqn:rearrangementuse}), we obtain:
\begin{align}
\widetilde{J} & \leq \widehat{J}
\end{align}
since $\alpha,\beta>0$. This contradicts the assumption that $\widehat{J}$ in Eq. (\ref{eqn:lossSSNNO}) is the globally optimal loss function. Thus,  the global minimizer of Eq. (\ref{eqopto}) will be an SSNNO with ordered variance of states. This completes the proof.
\end{proof}

Theorem \ref{lemmassnno} establishes that variance-ordering is achieved in SSNNO at the global minimizer by assigning variance-regularization weights in ascending order. 
However, during the training of SSNNO, as formulated in Problem \ref{SSNNO_problem}, the prediction accuracy of SSNNO relative to the SSNN may be influenced by the inherent trade-off between the SPE and model complexity, which is governed by the parameters $\alpha$ and $\beta$ in Eq. (\ref{eqloss}). The next theorem addresses this trade-off and provides an upper bound on the output prediction error for the SSNNO model.

\begin{theorem}[Variance regularization-SPE tradeoff]
\label{theoremtradeoff} 
The degradation in SPE of SSNNO (relative to SSNN) with the global minimizer of Eq. (\ref{eqopto}) for an appropriately chosen $\alpha$ and $\beta$ (Eq. (\ref{eqloss}))  is bounded by:
\begin{equation}
    \begin{aligned}
       J_{y}(\bm{\theta}^{*}) \leq \epsilon+\lambda
    \end{aligned}
\end{equation}
where $\lambda>0$ is a user-defined tolerance, and $\epsilon$ is a bound on the SPE of SSNN as given in Lemma 1 of \cite{bKK18}. 
\end{theorem}
\begin{proof}
    To prove the theorem, consider an SSNN with $d$ states and parameters $\bm{\theta}_{\text{SSNN}}^{*}$ obtained by minimizing the SPE ($J_{y}$ in Eq. (\ref{eqloss})).
Let $J_y(\bm{\theta}_{\text{SSNN}}^{*}) \leq \epsilon$ for an $\epsilon>0$ as given in \cite{bKK18}.  
Define,
\begin{equation}
\begin{aligned}
C_{1}=J_{v}(\bm{\theta}_{\text{SSNN}}^{*}), \hspace{0.5cm}C_{2}= J_{g}(\bm{\theta_{\text{SSNN}}^{*}})
\end{aligned}
\end{equation}
where $C_{1},C_{2}$ are the objective $J_{v}$ and $J_{g}$ as in Eq. (\ref{eqloss}) computed for $\bm{\theta}_{\text{SSNN}}^{*}$.  Choose $0<\alpha \leq \frac{\lambda-\beta C_{2}}{C_{1}}$ and $0<\beta< \frac{\lambda}{C_{2}}$ for a user-specified $\lambda>0$. Then, Eq. (\ref{eqloss}) gives:
\begin{equation}
\begin{aligned}
J(\bm{\theta}_{\text{SSNN}}^{*}) & = J_{y}(\bm{\theta}_{\text{SSNN}}^{*}) + \alpha J_{v}(\bm{\theta}_{\text{SSNN}}^{*}) + \beta J_{g}(\bm{\theta}_{\text{SSNN}}^{*}) \\ & \leq \epsilon + \alpha C_{1} + \beta C_{2}  \leq  
 \epsilon + \lambda.
\end{aligned}
\end{equation}
Consider the SSNNO to be obtained by minimizing the loss function $J$ in Eq. (\ref{eqloss}), as in Problem 1. Since $\bm{\theta}_{\text{SSNN}}^{*}$ is only a feasible solution of Problem \ref{SSNNO_problem}, it follows that for the optimal solution $\bm{\theta}^{*}$ of Problem \ref{SSNNO_problem}, we will have:
\begin{equation}
\begin{aligned}
&J(\bm{\theta}^{*})  = J_{y}(\bm{\theta}^{*}) + \alpha J_{v}(\bm{\theta}^{*}) + \beta J_{g}(\bm{\theta}^{*}) \leq J(\bm{\theta}_{\text{SSNN}}^{*})\leq \epsilon + \lambda  \\
&\implies J_{y}(\bm{\theta}^{*})  \leq \epsilon + \lambda, \hspace{0.7cm}\because~\alpha J_{v}(\bm{\theta}^{*}) + \beta J_{g}(\bm{\theta}^{*})\geq 0.
\end{aligned}
\end{equation}
Thus, for the SSNNO, the SPE for training data will not increase by more than $\lambda$ compared to the SSNN. 
\end{proof}

\subsubsection{A practical method to obtain SSNNO with guaranteed ordered variances for states}
\label{sec_practicalmethod}
 The proof of Theorems \ref{lemmassnno} and \ref{theoremtradeoff} are  based on the global minimizer $\bm{\theta}^{*}$. Although several recent works have proposed effective numerical methods for handling nonconvex optimization problems \cite{bFM25,bAB25},  conventional NN training techniques do not guarantee convergence to the global solution, in which case the predicted states may not satisfy the variance-ordering condition in Eq. (\ref{eqvarord}). Let $\bm{\theta}^{\mathrm{local}}$ denote a local minimizer for Eq. (\ref{eqopto}), with the corresponding loss function $\widehat{J}(\bm{\theta}^{\mathrm{local}})$ as in Eq. (\ref{eqn:lossSSNNO}). Similarly, let  $\widetilde{\bm{\theta}}^{\mathrm{local}}$ contain the parameters of the P-SSNNO obtained using the permutation $\mathbf{T}$ constructed as in Eq. (\ref{eqpermat}), for which the loss function be $\widetilde{J}(\widetilde{\bm{\theta}}^{\mathrm{local}})$ as given in Eq. (\ref{eqn:lossP-SSNNO}). Subsequently, following the same steps as in the proof of Theorem \ref{lemmassnno} yields:
 \begin{equation}
 \label{eqtheta_[ocal]}
     \widetilde{J}(\widetilde{\bm{\theta}}^{\mathrm{local}}) \leq \widehat{J}(\bm{\theta}^{\mathrm{local}}).
 \end{equation}
 Further, any $\bm{\theta}^{\mathrm{local}}$ with ordered state variance results in $\widetilde{J}(\widetilde{\bm{\theta}}^{\mathrm{local}}) = \widehat{J}(\bm{\theta}^{\mathrm{local}}),$ since $\widetilde{\bm{\theta}}^{\mathrm{local}} =\bm{\theta}^{\mathrm{local}},$ by construction. 
Building upon this,  a practical algorithm for obtaining SSNNO with guaranteed variance-ordering is presented below:
 \begin{algorithm}[H]
 	\begin{algorithmic}[1]	
	\STATE Given $\mathbf{U},$ $\mathbf{Y}$ in Eq. (\ref{equy})		
\STATE Select  $d, \delta,$ $\alpha,\beta$
  \STATE Initialize $\bm{\theta}$
		\STATE  Compute $\bm{\theta}^{\mathrm{local}}$ by solving optimization problem in Eq. (\ref{eqopto})
      \STATE  Construct a permutation matrix $\mathbf{T}$, as in Eq. (\ref{eqpermat}).
      \STATE Using $\mathbf{T}$, construct  $\widetilde{\bm{\theta}}^{\mathrm{local}}$ using Eqs. (\ref{eq_af1})-(\ref{eq_gtildei}).
      \STATE Compute $ \widehat{J}(\bm{\theta}^{\mathrm{local}})$ and $\widetilde{J}(\widetilde{\bm{\theta}}^{\mathrm{local}}) $ using Eq. (\ref{eqloss}) 
       \WHILE  {$\widehat{J}(\bm{\theta}^{\mathrm{local}})- \widetilde{J}(\widetilde{\bm{\theta}}^{\mathrm{local}}) > 0 $}  %
        \STATE  Initialize $\bm{\theta}=\widetilde{\bm{\theta}}^{\mathrm{local}}$
        \STATE Go to Step 4
       \ENDWHILE
        \STATE Return $\widetilde{\bm{\theta}}^{\mathrm{local}}$
	\end{algorithmic}
	\caption{Practical method for SSNNO training}
\end{algorithm}

It is important to note that, since the permuted parameter $\widetilde{\bm{\theta}}^{\mathrm{local}}$ leads to a non-increasing loss sequence as stated in Eq. (\ref{eqtheta_[ocal]}), Algorithm 1 converges to a local optimum that satisfies the variance-ordering condition.
%, or to the global optimum.

 The  SSNNO model 
can be used for designing data-driven state feedback-based control schemes such as MPC, adaptive control, etc. The next section presents a numerical implementation of the SSNNO on a CSTR example, which includes the application of the SSNNO in MPC.

%\newpage
\section{Simulation Results}
\label{secsim}
The proposed SSNNO is illustrated on a {continuous stirred tank reactor} (CSTR) system defined by the state equation in discrete time:
\begin{equation}
    \mathbf{x}_{k+1}=\mathbf{x}_{k}+ \int_{kT}^{(k+1)T} \mathbf{f}_{c} (\mathbf{x}(t),\mathbf{u}(t)) dt 
    \label{eqcstrst}
\end{equation}
where $T$ is the sampling period, $k$ is the discrete time instant, $t$ denotes continuous time, $\mathbf{f}_{c}$ is the state function of CSTR in continuous time \cite{bSJ01}: 
\begin{equation}
    \mathbf{f}_{c}=\left[\begin{matrix} -x_{1}+D_{a}(1-x_{1})e^{x_2} \\ x_{2}+B\hspace{.1cm}D_{a} (1-x_{1})e^{x_2}-D_{b}(x_{2}-u) \end{matrix}\right]
    \label{eqcstrstfc}
\end{equation}
where $x_{1}$, $x_2$ are the reactant conversion and reactor temperature, and $u$ is the reactor jacket temperature. Values of model parameters are: $B=22.0,D_{a}=0.082,D_{b}=3.0.$ 
The temperature of the reactor is the measured output:
\begin{equation}
  y_{k}=x_{2_k}+v_{k}
  \label{eqcstrop}
\end{equation}
where $v_{k}$ is the measurement noise, which is considered as Gaussian white noise with mean zero and standard deviation $0.05$.
It is noted that the CSTR system, with the true order $n=2$, exhibits severe gain nonlinearity. The gain between the concentration and jacket temperature exhibits an 80-fold change between low and high reactor temperature conditions \cite{bSJ01}. 
The input-output data is generated by a forward simulation of Eqs. (\ref{eqcstrst})-(\ref{eqcstrop}) over ${N}_s =900$ instants with the sampling period ${T}=1 \hspace{.1cm} second,$ initial condition $\mathbf{x}_{0}=[\begin{matrix} 0 ~~ 0 \end{matrix}]^\top.$ 
The control input $u$ is a multi-level pseudo-random signal in the range $[-0.6,0]$ with 45 step perturbations in the training period. This choice enables the input to be sufficiently rich, though a formal definition and quantification of the persistency of excitation 
for identification of nonlinear systems is an open problem.
Recently, persistency of excitation criteria for locally reachable nonlinear systems have been studied in \cite{bMA23}.  
\par {The first 500 samples of the dataset, denoted as $\mathbf{U}_{tr},\mathbf{Y}_{tr}$,  are used
for training the SSNNO, while the remaining samples denoted as $\mathbf{U}_{ts},\mathbf{Y}_{ts}$ are used for
 testing}.
%The SSNNO is trained using the training data where the loss function is chosen as in Eq. (\ref{eqloss}) with $\mathbf{U}=\mathbf{U}_{tr},$ $\mathbf{Y}=\mathbf{Y}_{tr}.$
The activation functions for the hidden layers are chosen as {hyperbolic tangent} (tanh) and the output layers as linear. The following parameter choices are made: ${d}=3,{L}=2,$ $l_{1}=3,$ ${H}=2,$  $h_{1}=3$, $\alpha=0.0025,\beta=0.25,\mathbf{W}=diag(1,2,3)$.  
The unconstrained optimization problem in Eq. (\ref{eqopto}) is solved for $\bm{\theta}$ using the Quasi-Newton method \cite{bNS06}. 
Fig. \ref{figcstr3}(a),(b) compare the system response with SSNNO with $d=3$ for the training and testing inputs. 
% \begin{table}[h!]
% \centering
% \caption{Performance comparison for fourth-order model.}
% \begin{tabular}{c c c   } 
%  \hline
% Performance measure & SSNNO & SSNN   \\ [0.5ex] 
%  \hline\hline
%   $V_{x_{1}}$ & 0.1372 &  0.0263  \\ 
%  $V_{x_{2}}$ & 0.0002 & 0.0047  \\
%  $V_{x_{3}}$ & 0.0000 & 0.0446 \\
%  $V_{x_{4}}$ & 0.0000 & 0.0431 \\
%  $\text{MSE}_{tr}$ & 0.0026 & 0.0025 \\
%  $\text{MSE}_{ts}$ & 0.0046 & 0.0049 \\
%  [1ex] 
%  \hline
% \end{tabular}
% \label{table:3}
% \end{table}

\begin{figure*}
	
 	\begin{center}
 		\includegraphics [scale=.35] {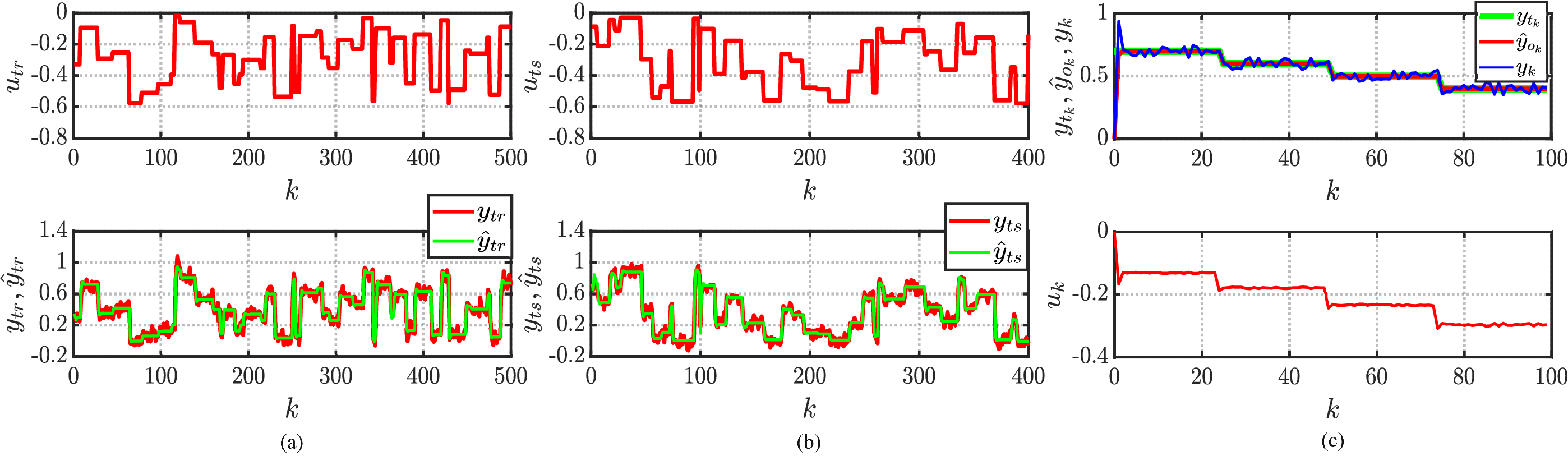}
 		\caption{{ {Response of SSNNO for CSTR with white noise: (a) Training, (b) Testing, (c) MPC response.}}}
 		\label{figcstr3}		
 	\end{center}
 \end{figure*}
 Table \ref{table:1}  shows the variances of the state variables with SSNNO and SSNN for $d=3$ where the initial parameters for the optimization problem Eq. (\ref{eqopto}) are chosen as uniform random variables.  It should be noted that the variances of the three states in SSNN are not ordered. Moreover, the variances are nonzero for all three state variables. On the other hand, the variances of the SSNNO states are ordered (despite using a local minimizer), and the variance of the third state variable is deemed insignificant, implying that the SSNNO is able to effectively determine an appropriate model order.  Further, Table \ref{table:1} shows the mean squared error ($\text{MSE}=\frac{1}{{N}}\text{SPE}$) between the given and predicted output for the training and testing data, denoted by $\text{MSE}_{tr} $ and $ \text{MSE}_{ts},$ respectively. 
 \begin{table}[h!]
\centering
\caption{Performance comparison of SSNNO and SSNN.}
\begin{tabular}{|c|c|c|c|c|} 
 \hline
\thead{Performance\\ measure} & SSNN & SSNNO & \thead{R-SSNNO\\($\delta=0.0005$)} & \thead{R-SSNNO\\($\delta=0.001$)}  \\ [0.5ex] 
 \hline 
 $V_{x_{1}}$  &  0.0029 & 0.1353  &  0.1368  & 0.1281 \\ 
$V_{x_{2}}$  & 0.0010 & 0.0006  &  0.0006 &  - \\
 $V_{x_{3}}$  & 0.0264 & 0.0001 & -  &  - \\
 $\text{MSE}_{tr}$  & 0.0027 & 0.0026   & 0.0028  &   0.0028\\
 $\text{MSE}_{ts}$  & 0.0025 & 0.0025  & 0.0026  &   0.0027\\
 [1ex] 
 \hline
\end{tabular}

\label{table:1}
\end{table}
%In the case of SSNNO with $d=4$, similar comparisons are shown in Table \ref{table:3}. It is noted that SSNNO achieves ordering of the states, with the last two state variables being insignificant. This demonstrates that the reduced model order determination in SSNNO does not depend on the user-specified model order $d$. 
% From Table \ref{table:1}, it can be observed that the output prediction errors with  SSNNO and SSNN are almost the same. 
%The extent of the model order reduction critically depends on the user-specified parameter $\delta$. It is clear that a choice of $\delta=0.0003$ would yield an SSNNO-R with only one state. 
Table \ref{table:1} also presents the results obtained using R-SSNNO models for $\delta = 0.0005$, which results in a second-order model, and for $\delta = 0.001$, which yields a first-order model.
 The output prediction errors obtained with  R-SSNNO models are almost the same as the third-order SSNNO model, indicating that a second-order (or first-order) 
 R-SSNNO model is a reasonable choice. 

\par To demonstrate the practical method outlined in Algorithm 1, the SSNNO optimization problem Eq. (\ref{eqopto}) is solved using an alternative initial parameter, which leads to a local optimum where the states are not ordered in variance. The results, presented in Table \ref{table:2} compare the outcomes obtained using SSNNO (for the local optimum), SSNNO-2 (using Algorithm 1), and SSNN. In particular, the $\tilde{\theta}^{\text{local}}$ obtained from Algorithm-2 was used as an initial guess in Step 3 and the local optimizer was re-run to obtain an updated $\tilde{\theta}^{\text{local}}$. From Table \ref{table:2}, it can be observed that the SSNNO with the local optimum corresponding to the specific initial parameter does not result in variance-ordering for which the variance and MSE values are similar to SSNN. Whereas, with the practical method SSNNO-2 is able to achieve variance-ordering for this initial parameter with a satisfactory prediction accuracy. 
\begin{table}[h!]
\centering
\caption{Illustrating the practical method (Algorithm 1)}.
\begin{tabular}{|c|c|c|c|} 
 \hline
\thead{Performance\\ measure} & SSNN & SSNNO & SSNNO-2    \\ [0.5ex] 
 \hline 
 $V_{x_{1}}$ &  0.0012 & 0.0014 &  0.1313    \\ 
$V_{x_{2}}$ &0.0914 & 0.0888 & 0.0022  \\
 $V_{x_{3}}$& 0.0555 & 0.0516 & 0.0003  \\
 $\text{MSE}_{tr}$ & 0.0502 & 0.0494  & 0.0026  \\
 $\text{MSE}_{ts}$ & 0.0595 & 0.0568 &  0.0027   \\
 [1ex] 
 \hline
\end{tabular}

\label{table:2}
\end{table}
\par Table \ref{table:3} presents the results of a Monte Carlo simulation conducted under six different noise levels on the training and testing output data, with standard deviations ranging from $0.01$ to $0.06$ in increments of $0.01$. The mean and variance of the quantities reported in Table \ref{table:3} were computed by training SSNNO on the six datasets described above using five different initial parameters. The solution corresponding to the minimum training error among these runs for each dataset was used in the calculation of the mean and variance.  From Table \ref{table:3}, it can be observed that ordering is achieved in SSNNO without compromising the SPE performance. %In the simulation case study, variance ordering is achieved by directly solving the optimization problem in Eq. (\ref{eqopto}). Therefore, an opportunity to use the practical method discussed in the previous section did not arise. 

\begin{table}[h!]
 \centering
 \caption{{Monte Carlo simulation with different noise levels}.}
 \begin{tabular}{|l|c|c|c|c|}
 \hline
 \multicolumn{1}{|c|}{Perf.} & \multicolumn{2}{c|}{SSNNO} & \multicolumn{2}{c|}{SSNN}\\
 \cline{2-5}
 \multicolumn{1}{|c|}{measure} & Mean & Variance & Mean & Variance \\
 \hline
 $V_{x_1}$   & 0.1407    & $7.8\times 10^{-7}$ &0.0224& $3.9 \times 10^{-4}$\\
 $V_{x_2}$&   0.00013  & $5.4\times 10^{-8}$&0.0094&  $6.9 \times 10^{-5}$   \\
 $V_{x_3}$ &0.00010 & $8.0 \times 10^{-9}$&0.0102&$6.4 \times 10^{-5}$ \\
  $\text{MSE}_{tr}$   & 0.0014 & $1.4\times 10^{-6}$&0.0014&$1.5 \times 10^{-6}$ \\
  $\text{MSE}_{ts}$&   0.0026  & $1.6 \times 10^{-6}$&0.0015& $1.4 \times 10^{-6}$ \\
 \hline
 \end{tabular}
 \label{table:3}
 \end{table}

 \subsection{EKF-based state space MPC design: CSTR Example}
 This section illustrates the use of the SSNNO as the prediction model for the MPC of the CSTR system. The cost function for the MPC is chosen as a quadratic function of state and control input \cite{bDM21}.
 \begin{equation}
 \label{eqMPCJk}
 \begin{aligned}
     J_k=& \sum_{i=k+1}^{k+{N_p}}[\mathbf{x}_{\text{ref}_{i|k}}-\mathbf{x}_{_{i|k}}]^{\top}\mathbf{Q}[\mathbf{x}_{\text{ref}_{i|k}}-\mathbf{x}_{_{i|k}}] + \sum_{i=k}^{k+{N_p}-1} [\mathbf{u}_{\text{ref}_{i|k}}-\mathbf{u}_{i|k}]^{\top}\mathbf{R}[\mathbf{u}_{\text{ref}_{i|k}}-\mathbf{u}_{i|k}] 
    \end{aligned} 
 \end{equation}
 where $\mathbf{x}_{_{i|k}},\mathbf{u}_{i|k}$ denotes the predicted state and control input at time instant $i$ but computed at time instant $k$, and $\mathbf{x}_{\text{ref}_{i|k}},\mathbf{u}_{\text{ref}_{i|k}}$ denotes the reference state and control inputs. The weighting matrices for state and control inputs are chosen as $\mathbf{Q}=diag(0.5,1)$, $R=0.5$, and the MPC horizon is chosen as $N_{p}=5.$  Control constraints are imposed as $u_{min}\leq \mathbf{u}_{k} \leq u_{max}$ with $u_{min}=-1$ and $u_{max}=0.$
The output targets for the reactor temperature are chosen as $\mathbf{y}_{\text{t}}=0.7$ for the first quarter (first 25 instants), which is then reduced by 0.1 in each successive quarter. 
The state reference $\mathbf{x}_{\text{ref}}$ and control reference $\mathbf{u}_{\text{ref}}$ used in the MPC cost function  are computed by solving the steady-state R-SSNNO model (with order $s=2$) as follows:
\begin{equation}
\label{eqref}
\left[\begin{matrix}  \mathbf{y}_{\text{t}}-\textbf{g}_{s}(\mathbf{x}_{\text{ref}}) \\  \mathbf{x}_{\text{ref}}-\mathbf{f}_{s}(\mathbf{x}_{\text{ref}},\mathbf{u}_{\text{ref}})    \end{matrix}\right]
=\mathbf{0}
    \end{equation}
for each of the reactor temperature targets. The MPC scheme uses states estimated using an EKF as in \cite{bLY13} with disturbance and noise covariance matrices $\mathbf{Q}_{\text{c}}=diag(0.1,0.2)$, and $\mathbf{R}_{\text{c}}=0.1$, respectively. The resultant scheme is denoted as SSNNO-EKF-MPC, for which 
 Fig. \ref{figcstr3}(c) shows the output response and control input. Here  $\widehat{y}_{k}$ denotes the output predicted by the SSNNO model, and ${y}_{k}$ is the output with the first principle model (in Eq. (\ref{eqcstrop})) for the MPC control input shown in Fig. \ref{figcstr3}(c) bottom panel. 
 From Fig. \ref{figcstr3}(c), it can be observed that
the output of the CSTR system follows the reference value with the proposed SSNNO-EKF-MPC scheme. 
There is an initial overshoot in the output with the proposed scheme, which is due to the plant-model mismatch, which is corrected using an EKF. The simulation codes are available on GitHub\footnote{https://github.com/MIDHUNTA30/SSNNO-MATLAB}.
\section{Conclusions}
A state-space neural network with ordered variance (SSNNO) is proposed as an effective method for identifying the model order directly from training data. This approach leads to the identification of a nonlinear state-space model in which the state variables are arranged in descending order of their variances, ensuring that the most significant dynamics are captured first. Additionally, the method facilitates the derivation of a reduced-order SSNNO model, which retains the essential dynamics in reduced-order manifold while reducing the model complexity. %Thus, the variance plays the role of eigenvalues for eigen-decomposition based model order reduction techniques used in linear systems \cite{bED66}. 
%The R-SSNNO model is capable of predicting system outputs with sufficient accuracy, making it a practical tool for control and estimation tasks. 
Theoretical results are presented on variance-ordering and output prediction error bound with SSNNO. The effectiveness of the proposed approach is demonstrated through simulations on a continuous stirred tank reactor system. 
Future research directions include extending the SSNNO framework to provide theoretical guarantees on properties such as controllability, and stability, %persistency of excitation, 
which would 
enhance the utility
%further enhance the reliability and explainability 
of the approach in practical control applications.

% \section*{Acknowledgments}
% This research was supported by the Science and Engineering Research Board, Department of Science and Technology  India, through grant number CRG/2022/002587.

\end{document}